\documentstyle[12pt]{article}
\begin{document}
\oddsidemargin 0mm  \evensidemargin 0mm  \topmargin -25.4mm
\headheight 42mm    \headsep 0mm
\textheight 212mm   \textwidth 152mm
\footheight 42mm    \footskip 15mm

\newcommand{\bq}{\begin{equation}}
\newcommand{\eq}{\end{equation}}
\newcommand{\bqa}{\begin{eqnarray}}
\newcommand{\eqa}{\end{eqnarray}}
\newcommand{\nl}{\nonumber\\}
\newcommand{\p}[1]{{\scriptstyle{\,(#1)}}}
\newcommand{\eqn}[1]{eq.~(\ref{#1})}
\newcommand{\skl}[1]{\vspace{#1\baselineskip}}
\newcommand{\order}[1]{{\cal O}\left(\alpha^{#1}\right)}
\newcommand{\bov}[1]{$^{\mbox{{\small #1}}}$}
\newcommand{\gev}{\mbox{GeV}}

\pagestyle{empty}
\begin{flushright}
preprint INLO-PUB-6/94\\preprint NIKHEF-H/94-20\\
hep-ph/9405398
\end{flushright}
\skl{1}
\begin{center}
{\Large {\bf Initial-state QED corrections\\ to four-fermion production
     in $e^+e^-$ collisions\\ at LEP200 and beyond} \bov{a}}\\
\skl{1}
{\bf F.A.~Berends \bov{b} and R.~Pittau \bov{c}}\\
Instituut-Lorentz, University of Leiden, the Netherlands\\
\skl{1}
{\bf Ronald Kleiss\bov{d}}\\
NIKHEF-H, Amsterdam, the Netherlands\\
\skl{2}
{\bf Abstract}
\end{center}
The implementation of QED initial-state radiative
corrections in the process of four-fermion production at LEP200
and higher-energy $e^+e^-$ colliders is discussed.
Because of the presence of charged-current processes,
this is a nontrivial problem, and we compare our approach
with other existing treatments. We describe the Monte Carlo algorithm
used for the generation of 4-fermion events with photon bremsstrahlung.
Comparison between our event generator and semi-numerical calculations
are presented, as well as predictions for $W$ and $Z$ pair related
cross sections.\\
\skl{2}
\begin{flushleft}
--------------------------------------------------------\\
\begin{small}
\bov{a} Research supported in part by the EU under
contract no. CHRX-CT-0004.\\
\bov{b} e-mail: berends@rulgm0.leidenuniv.nl\\
\bov{c} e-mail: pittau@rulgm0.leidenuniv.nl\\
\bov{d} e-mail: t30@nikhefh.nikhef.nl
\end{small}\end{flushleft}

\newpage
\pagestyle{plain}
\setcounter{page}{1}

\section{Introduction}
In this letter we discuss the implementation of initial-state QED radiation
(ISR) in the reaction
\bq
e^+ e^-\;\;\;\rightarrow\;\;\;4\;\mbox{fermions}\;\;.
\label{4ferm}
\eq
We shall discuss our method, present a variety of results for the
energy regime of LEP200 and slightly above, and, where possible,
make comparisons with other existing results. The purpose of this study is
to provide an efficient and flexible calculational tool, in the form of
a Monte Carlo event generator, {\tt EXCALIBUR},
that incorporates the expected dominant
physics effects. Therefore, this tool should be useful for all kinds of
physics studies at LEP200 and beyond. We start by outlining our method.

\section{The method}
In a recent paper \cite{1}, we studied all electroweak
(EW) $e^+e^-$ processes
leading to a final state of four effectively massless fermions,
without regard for the effects of photon emission.
In the case of four quarks, one would like to add the concomitant QCD
production channel, and also the production of a quark pair and two
gluons, since both different final states will appear as jets. This
QCD-extended treatment is now also availabe \cite{2}.

The motivation for our considering 4-fermion final states is that the
single or pairwise production of vectorbosons, at LEP200 energies or
beyond, typically manifest themselves as 4-fermion production.
A priori, one does not know the relative sizes of the production
{\it via\/} an intermediate step of vector boson production (the
`signal'), or {\it via\/} direct production (the `background'), and,
moreover, these two alternatives interfere; for the interpretation
of future experiments this knowledge is requisite. The results of
\cite{1} show that, for a number of final states, background effects
can be as large as radiative-correction effects.

In \cite{1}, a general procedure was developed in which any
4-fermion final state can be chosen, and (weighted) Monte Carlo
events efficiently generated. In this way, every possible distribution
or cross section in these processes becomes calculable. The only
restriction in \cite{1} is that charged particles in the final state
be, experimentally, visible and distinguishable, {\it i.e.\/}
they must be produced with sufficient energy, and
at nonnegligible angle to the beams or to
each other. Under this condition, one avoids collinear singularities:
therefore, the fermions can be treated as massless, which considerably
accelerates the calculation \footnote{Note that signals with an
intermediate Higgs therefore fall outside of our scope.}. When one
considers only collinearly {\em convergent\/} diagrams, we may even
omit our visibility requirements. Amongst others, this case occurs
for the $W$ pair signal reaction
\bq
e^+ e^-\;\;\;\to\;\;\;W^+W^-\;\;\;\to\;\;\;4\;\;\mbox{fermions}\;.
\label{viaww}
\eq
In the calculation of the Born diagrams that are needed in the event
generator, a number of input parameters can be chosen independently,
to wit, the electroweak couplings parametrized by a (running) $\alpha$
and $\sin^2\theta_w$, the boson masses $m_W$ and $m_Z$, and the
corresponding total widths $\Gamma_W$ and $\Gamma_Z$. For instance,
one has the freedom to choose $\cos\theta_w$ different from
$m_W/m_Z$, and to take arbitrary values for the widths, which may also
be assigned an energy dependence. This convention, in which $\alpha$
occurs as an overall factor in the matrix element, automatically ensures
the unitarity cancellations, so that the value of $\alpha$ may be
changed at will.

Although the event generator of \cite{1} is suitable for many
signal-versus-background studies, its usefulness is increased considerably
when the most important radiative correction effects are incorporated.
Let us discuss how this can be achieved in a practical way.

\section{Inclusion of radiative corrections}
To start, we focus on the reaction
\bq
e^+ e^-\;\;\;\to\;\;\;W^+W^-\;\;,
\label{wpairs}
\eq
with stable $W$'s ($\Gamma_W=0$), and on the reaction in
\eqn{viaww}, with decaying $W$'s ($\Gamma_W>0$). The
radiative-correction problem is much more involved than for the, by
now familiar, case (see, {\it e.g.\/} \cite{3})
\bq
e^+ e^-\;\;\;\to\;\;\;Z^0\;\;\;\to\;\;\;\mu^+ \mu^-\;\;.
\label{zmumu}
\eq
In the latter reaction, photonic and weak corrections can be
separated. Thus, the weak one-loop corrections lead to a modification
of the Born cross section into a `dressed' Born cross section.
To this latter, the sizeable QED initial-state corrections can be applied
by means of the structure-function method. These QED corrections
incorporate $\order{}$ and $\order{2}$ leading-log (LL)
and subleading terms, while the leading higher-order soft-photon
contributions are implemented by exponentiation.

Despite many attempts, a simple dressed Born cross section for the
reaction of \eqn{wpairs} has not be found, so that the above
procedure for the $Z$ cannot be applied: the weak corrections do not
decouple from the the QED ones in $W$ pair production. A cognate
difficulty is that a division of photon emission into initial- and
final-state radiation is meaningful for reaction (\ref{zmumu}) but not
for (\ref{wpairs}), since the two sets of Feynman diagrams are not
separately gauge-invariant. A priori, then, it is meaningless to
consider those for initial-state radiation in isolation. An elegant
trick has been proposed in \cite{4} to
introduce a gauge-invariant definition of initial- and final-state
terms, based on adding and subtracting extra terms in the radiative
matrix element. Although solving, in a sense, the problem of
gauge invariance, there is as yet no proof that the initial-state
radiation terms thus obtained yield a quantitatively good description.
Since the LL terms in the cross section are anyhow gauge-invariant,
the method of \cite{4} is correct for these terms, but the subleading ISR
terms now contain some arbitrariness.

In order to facilitate the discussion that follows, we find it useful
to summarize in table 1 what has been done in the literature, and
what some authors hope to achieve.
We also indicate the position of the present paper in this welter of
possibilities.
\begin{table}[ft]
\begin{center}\begin{tabular}{|c|c|c|c|} \hline \hline
 & \multicolumn{3}{|c|}{Final state} \\ \hline
RC treatment       &$W$'s            &4 fermions &4 fermions\\
                   &on-shell         &from $W'$s &all diags.\\ \hline
$\order{}$ EW      &                 &           &          \\
+ 1 soft $\gamma$  &\cite{5}-\cite{8}&\cite{13}  &\cite{13} \\ \hline
$\order{}$ EW      &                 &           &          \\
+ 1 hard $\gamma$  &\cite{9}         &\cite{13}  &\cite{13} \\ \hline
same, but          &                 &           &          \\
event generator    &\cite{10,10a}    &           &          \\ \hline
LL QED ISR +       &                 &           &          \\
exponentiation     &\cite{11}        &\cite{4}   &          \\ \hline
Full EW $\order{}$ &                 &           &          \\
+ high.ord. LL ISR &\cite{9,12}      &           &          \\ \hline
LL QED ISR + exp.  &                 &\cite{14}, and &      \\
event generator    &                 &this paper &this paper\\ \hline \hline
\end{tabular}\end{center}
\caption[.]{A summary of references to radiative-correction
studies of $W$ pair or 4-fermion production. Either electroweak
or LL QED ISR have been considered. Event-generator approaches
are mentioned explicitly. All references contain numerical
results, except ref. \cite{13} which devises a strategy for
the most complete treatment.}
\end{table}

For stable $W$'s, complete one-loop EW effects have been calculated
in a number of papers \cite{5,6,7,8}. The results of
refs. \cite{7} and \cite{8} are in perfect agreement.
To obtain the complete $\order{}$ correction, the effect of
emission of a single hard photon has to be included \cite{9}.
An event generator based on this calculation also exists \cite{10},
whereas recently another event generator has been constructed where, for
collinear photons, $\order{2}$ QED corrections are also considered \cite{10a}.

In order to asses the importance of higher-order photonic effects,
a LL calculation up to $\order{2}$, including exponentiated
soft-photon effects, was already carried out in an early for the
total cross section and for
an angular distribution \cite{11}. Somewhat later, the $\order{}$
term of such a QED LL treatment was replaced by a full $\order{}$ EW
calculation \cite{9,12}. The difference between these two approaches amounts to
less than 1\% in the LEP200 energy range.

For decaying $W$'s, the radiative corrections are far less known.
The full $\order{}$ EW corrections have not yet been
computed. Questions of strategy, and theoretical issues like gauge invariance,
have been addressed in the literature, which may form the basis for
a conclusive calculation \cite{13}.
What {\em has\/} been done, are ISR calculations to the 4-fermion final
state produced by the $WW$ signal. In ref. \cite{4}, the LL plus
exponentiated soft-photon contributions are evaluated semi-analytically.
For the total cross section, a 3-dimensional integral has to be done
numerically, after a number of analytic integrals have been performed.
Thus, only very specific distributions can be obtained. The
influence of subleading terms in the ISR are studied, but, since the
above-mentioned trick has been used in \cite{4} in order to
give a gauge-invariant definition of ISR, the {\em quantitative\/}
meaning of precisely these terms is obscure until a full calculation,
including final-state radiation (and interference) has been achieved.
Another approach uses LL QED corrections to the signal diagrams,
leading to an event generator for the three signal diagrams \cite{14}.\\

In the present paper, we want to be able to study {\em any\/}
experimental distribution \footnote{We except the transverse-momentum
distribution of the bremsstrahlung.} and all background effects,
and we want to incorporate the dominant radiative correction effects.
This leads to an event generator that can handle all diagrams leading to
a specified 4-fermion final state (with, of course, the option of
a restriction to the signal diagrams), and that incorporates the LL
$\order{}$ and $\order{2}$ ISR, with exponentiation of
the remaining soft-photon effects. We shall now discuss how the
initial-state radiation effects are incorporated in our
Monte Carlo event generator \cite{1}.

\section{Generating initial-state radiation}
In order to upgrade the event generator of ref. \cite{1} with
QED ISR, the following description of the radiation is used.
Each of the incoming fermions is assumed to have its energy degraded
by an amount of bremsstrahlung. Under the assumption
that the bremsstrahlung
photons are emitted parallel to the radiating beam, the energy
distribution of the fermion after radiation is described by
the `structure' function
\bqa
\Phi(x) & = & {\exp\left(-\beta\gamma_E+3\alpha L/4\pi\right)\over
               \Gamma(1+\beta)}\beta(1-x)^{\beta-1}
         -{\alpha\over2\pi}(1+x)L\nl
 & & -{\alpha^2\over8\pi^2}\left[{1+3x^2\over1-x}\log x
 +4(1+x)\log(1-x)+5+x\right]L^2\;\;,\nl
\beta   & = & {\alpha\over\pi}(L-1)\;\;\;,\;\;\;
L = \log\left({Q^2\over m_e^2}\right)\;\;,
\label{struct}
\eqa
where $x$ is the fermion's energy in units of the beam energy,
$m_e$ is the electron mass, $\gamma_E$ is Euler's constant,
and $Q^2$ is some appropriate energy scale. This is a LL
$\order{2}$
structure function with exponentiated soft-photon effects.
Subleading logarithmic terms are not considered, since one needs for this
the non leading logarithmic terms from the one loop EW corrections to
the Born cross section of reaction in \eqn{4ferm}.
Another important point is the choice of $Q^2$.
Formally, a change in $Q^2$ is a subleading effect, but its precise
value is of course a matter of numerical concern. We use $Q^2=s$, the
total energy squared, which is known to be acceptable \cite{12}.
It is a unique advantage of the Monte Carlo
approach that, if we wish, the $Q^2$ can even be determined on
an event-by-event basis.
Finally, the structure functions $\Phi(x_1)$
for the incoming $e^+$ (with original momentum $p_1$, degraded to
$x_1p_1$) and $\Phi(x_2)$ for the incoming $e^-$ (with momentum
$p_2$, degraded to $x_2p_2$), can be convoluted: with
\bq
s' \equiv (x_1p_1+x_2p_2)^2 = x_1x_2s\;\;,
\eq
we arrive at the `flux function'
\bq
G(s'/s) = \int\limits_0^1\int\limits_0^1\;
dx_1\;dx_2\;\Phi(x_1)\;\Phi(x_2)\;\delta(x_1x_2-s'/s)
\eq
which enables one to write the total radiative cross section as
\bq
\sigma(s) =  \int\limits_0^1\;dz\;G(z)\sigma_0(zs)\;\;,
\eq
where $\sigma_0$ is the nonradiative cross section.
Incidentally, the form of $G(x)$ is, to given order in $\alpha$,
quite close to that of $\Phi(x)$, where $\alpha$ is replaced by
$2\alpha$ \cite{12}. In LL approximation the flux function $G_D$ of
ref. \cite{3} is related to the choice made in \eqn{struct}.
We have opted for the use
of structure functions rather than that of a flux function for the
following reason. Assuming that the four-momentum lost to
radiation is lightlike (as one usually does in problems such as this one),
the total {\em energy loss\/} of the beams must be equal,
in the flux-function formalism, to $(1-s'/s)\sqrt{s}/2$; for instance,
this identification is made in \cite{4}. The energy loss can, however,
be quite different in the structure-function formalism, where the
lost momentum is the sum of two lightlike vectors; and this is the
more realistic approach, since radiation from the two beams tends to
be contained in two narrow, back-to-back cones. Another way to
appreciate the difference is to note that, when all the lost momentum
is lumped into a single lightlike vector, events with small $s'$ will
{\em always\/} be boosted away from the lab frame, whereas in the
structure-function formalism they can easily be at rest in the lab frame.\\

We are now in a position to decribe the radiation algorithm.
To start, two values for $x_1$ and $x_2$
are generated, each with a probability
distribution
\bqa
\Phi_1(x) & = & {1\over\beta_1}(1-x)^{\beta_1-1}\;\;,\nl
\beta_1 & = & {\alpha\over\pi}\left(\log{Q^2_1\over m_e^2}-1\right)\;\;,
\eqa
where we have introduced yet another scale $Q^2_1$. We then compute
$x_1p_1$ and $x_2p_2$, so that the total momentum of the 4-fermion
system is determined. This allows us to move over to the centre-of-mass
frame of the 4-fermion system; here, the total energy is $\sqrt{s'}$
rather than $\sqrt{s}$, and, importantly, the beam directions are the
same as in the lab frame.
This is due to the assumption that the bremsstrahlung is strictly
collinear: relaxing this assumption would lead to, as yet nearly
insuperable, complications in the Monte Carlo.

We now generate a weighted Monte Carlo 4-fermion event by the procedure
described in \cite{1}. The resultant momenta are then boosted back to the
lab frame. The final event weight is then the product of the
`4-fermion weight' defined in \cite{1}, and the `radiation weight'
\bq
w_{\mbox{{\small rad}}} =
{\Phi(x_1)\over\Phi_1(x_1)}{\Phi(x_2)\over\Phi_1(x_2)}\;\;.
\eq
As usual, the Monte Carlo cross section and its estimated error are
then extracted from the mean and the variance of the weight
distribution. We want to stress that the adoption of a
structure function different from that of \eqn{struct} is
trivial in our approach, since it only entails modifying the definition
of $w_{\mbox{{\small rad}}}$: in this way, we can, for instance,
perform delicate checks with the results of ref. \cite{4}.

A few remarks are in order here. In the first place, note that
the use of two structure functions with terms up to, say
$\order{}$, is {\em not\/} equivalent to a calculation based on
a flux function to $\order{}$, since the product of the two
structure functions contains some $\order{2}$ terms. Of course,
with the above algorithm we may also settle for generating
only a single value for $z=x_1x_2$, and proceeding with
the generation of the 4-fermion final state using the reduced
energy $s'=zs$. Since, however, the Lorentz boost is then not determined,
we can only compute the total cross section in that case, and no
differential ones. Nevertheless, we have applied this procedure
in order to compare our event generator with the semi-analytical results
of ref. \cite{4}.

Secondly, we have to discuss the choice of $Q^2_1$. In principle, it may
be chosen at will, since the Monte Carlo cross section does, formally,
not depend on it. However, it is easily seen that
$w_{\mbox{{\small rad}}}$ will diverge as $x_1$ or $x_2$ approache
one, unless $Q^2_1\le Q^2$. We therefore must, in the generation of events,
choose $Q^2_1$ to be the minimum possible value for $Q^2$. If one
desires to use a fixed $Q^2$ scale (as we have done
in this paper) this poses no problem, and one
simply puts $Q^2_1=Q^2$; but for a study where $Q^2$ depends on the
particular event generated, some care must be taken.

Finally, it should be realized that the
various kinematical cuts described
above must also be boosted to the 4-fermion rest frame. For
invariant-mass cuts this is no problem, but the energy and angular cuts
require some attention. Given two values $x_1$ and $x_2$,
we know the relativisitc velocity of the Lorentz boost to the
4-fermion centre-of-mass frame (CMF):
\bq
\beta=(x_1-x_2)/(x_1+x_2)\;\;.
\eq
If the scattering angle $\theta$ of a (massless) particle in the lab frame
is restricted between
\bq
c_0\;<\;\cos\theta\;<\;c_1\;\;,
\eq
we may then
compute the bounds in the CMF on the CMF scattering angle $\theta'$:
\bq
{c_0-\beta\over1-c_0\beta}\;<\cos\theta'\;<
{c_1-\beta\over1-c_1\beta}\;\;.
\eq
Similarly, suppose its energy $E$ in the lab is restricted by
\bq
E\;>\;E_{\mbox{{\small min}}}\;\;.
\eq
Its energy $E'$ in the CMF now depends on both the energy and the
angle in the lab frame, so that a CMF energy cut is complicated;
we replace it by its lower bound (the minimum over all scattering
angles):
\bq
E'\;>\;{E_{\mbox{{\small min}}}\over\sqrt{1-\beta^2}}
\left(\vphantom{A^1_1}1-\max(c_0\beta,c_1\beta)\right)\;\;.
\eq
Since this cut is somewhat looser, some particles may end up with
an energy lower than $E_{\mbox{{\small min}}}$;
this means that an additional number of generated events has to
be rejected. It must be stressed that these cuts would become
impractically complicated if the bremsstrahlung were also to have
a transverse momentum component.

\section{Results and conclusions}
We shall now discuss a number of results from {\tt EXCALIBUR}.
In the first place, we have to establish agreement, where possible,
with the results of ref. \cite{4}. To this end, we must of course make sure
to use the same electroweak input parameters. In the structure functions
and the flux functions we use the low-energy value
\bq
\alpha_{\mbox{{\small str.f., flux}}} = (137.036)^{-1}\;\;.
\eq
In the rest of the calculation, we take the input parameters
such that important electroweak corrections to the reaction
(\ref{wpairs}) are effectively incorporated. In the notation of
ref. \cite{12}, the amplitude $M$ for reaction (\ref{wpairs}) can be
divided into two gauge-invariant parts:
\bq
M = {e^2\over2\sin^2\theta_w}M_I + e^2M_Q\;\;,
\eq
where $M_I$ is the purely $V-A$ part, and $M_Q$ a purely vector-like
part. The number $e^2=4\pi\alpha$ we choose such that the running value
of $\alpha$ at LEP200 energies is obtained, and $\sin^2\theta_w$ we fix by
\bq
{\alpha\over2\sin^2\theta_w} = {G_{\mu}m_W^2\over\pi\sqrt{2}}\;\;.
\eq
This gives us the parameters also used in \cite{4}. We also adopt the
values used there for the boson masses and widths. Numerically,
we then have
\bqa
& & \alpha = (127.29)^{-1}\;\;\;,\;\;\;
\sin^2\theta_w = 0.2325\;\;\;,\;\;\;
G_{\mu} = 1.16635\;10^{-5}\;\gev^{-2}\;\;\;,\nl
& & m_Z = 91.173\;\gev\;\;\;,\;\;\;\Gamma_Z = 2.4971\;\gev\;\;\;,\nl
& & m_W = 80.220\;\gev\;\;\;,\;\;\;\Gamma_W = 2.033\;\gev\;\;.
\eqa
We use, moreover, boson propagators with energy-dependent widths;
their denominators have the typical form
$s-m^2+is\Gamma/m$.
In \cite{12}, a discussion of an `improved Born approximation' is
given. In our approach, where we have both signal and background
effects, we can retain of this discussion only the above-mentioned
tuning of $\alpha$ and $\sin^2\theta_w$. In particular, the effects
of the Coulomb singularity in the $WW$ system \cite{15,16}, is
left out.

We are now ready to present some saliant results.
We have chosen the explicit process
\bq
e^+ e^-\;\;\to\;\;e^- \bar{\nu_e} u \bar{d} \;\;,
\eq
which contains the $WW$ pair production signal, as well
as a nonnegligible background.
In the first place,
we must reproduce the results of \cite{4,17} where we can. These are
the Born cross section $\sigma_0$; the
total cross section with ISR from ref. \cite{17}
and the average `energy loss',
which in \cite{4} is defined as ${\sqrt{s}\over 2}(1-x_1x_2)$, which we denote
by $\bar{\epsilon}$.
This is (see above) not the {\em real\/} energy loss
${\sqrt{s}\over 2}(2-x_1-x_2)$, denoted by $\epsilon$; but which of the two
quantities is actually the most relevant in the measurement of
the $W$ mass is, of course, determined by the prospective data analysis.

We have performed a number of different Monte Carlo studies, under
different strategies, which we now list.
\begin{itemize}
\item {\bf WW,f,1}: the $WW$ signal diagrams only, with the
  flux-function appraoch to $\order{}$; no cuts.
\item {\bf WW,s,1,a}: the $WW$ signal diagrams only, with structure
  functions that are simply the flux function of \cite{4}
  to $\order{}$, in which $2\alpha$ is replaced by $\alpha$; no cuts.
\item {\bf WW,s,1,b}: the $WW$ signal diagrams only, with the
  structure functions of \eqn{struct} where the last ($L^2$) terms
  are left out; no cuts.
\item {\bf WW,f,2}: the $WW$ signal diagrams only, with the
  complete $\order{2}$ flux function of ref. \cite{4} (which,
  unfortunately, is not given explicitly); no cuts.
\item {\bf WW,s,2}: the $WW$ signal diagrams only, with structure
  functions as given in \eqn{struct}; no cuts.
\item {\bf WW,cuts}: like the previous case, except that now
  we also impose the following cuts: \\
$E_{e^-,~u,~\bar{d}}>20~GeV$, $|\cos \theta_{e^-,~u,~\bar{d}}|<0.9$,
$|\cos \p{u\bar{d}}|<0.9$, $m\p{u\bar{d}}>10~GeV$.
\item {\bf all,cuts}: like the previous case, except that now also
  all the background Feynman diagrams are taken into account.
\end{itemize}
The various results are given in table 2, where the numbers taken over
from ref. \cite{4,17} are indicated by the subscript 4.
The cross sections are given in picobarns, and the energy losses
in GeV.

Similarly in table 3 results are given for Z-pair production. The choice of
the second energy was determined by the condition that the velocities of the
$Z'$s would be the same as for the $W'$s at $\sqrt{s}= 190$ GeV, when there
is no radiation present. The cuts  used are $E_\p{all~particles}>20~GeV$,
$|\cos \theta_{\p{all~particles}}|<0.9$, $m\p{e^+e^-}$ and $m\p{u\bar{u}}>$
10 GeV and $|\cos \p{u\bar{u}}|<$ 0.9.

In discussing the results, let us first notice that the cross sections
from the event generator and the semi-analytical approach \cite{4,17}
agree. For $\sigma_0$ it is clear when one takes the numbers .60078 and .67930
from \cite{17}, for $\sigma$ it follows from the flux function comparisons
in table 1. These comparisons agree within 1 standard deviation. For
$\bar{\epsilon}$ and $\bar{\epsilon}_4$ there are differences up to five
standard deviations. The error on the results for $\bar{\epsilon}_4$ of
ref. \cite{4} is however not available \cite{17}, so that no conclusion
can be drawn.

The structure function method differs slightly from the flux function method
as comparisons between the first and second rows of table 1 shows.
The $\order{2}$ results from {\tt EXCALIBUR} and ref. \cite{4} differ at
the 2\% level for the energy losses, but this may be due to the not specified
form of $\order{2}$ corrections in \cite{4}.

The inclusion of cuts  and the inclusion of more diagrams affect both cross
sections and energy losses.

Since the proposed direct reconstruction method for the $W$ mass suffers from a
 shift in $M_W$ due to the radiated energy \cite{18}, a precise knowledge
of $\bar{\epsilon}$ and $\epsilon$ is warranted. Our results show that the
precise treatment of ISR, the choice of cuts and the neglect of diagrams all
affect the energy losses. The first effect was also found in ref. \cite{4},
where also the influence of the Coulomb singularity is discussed.
The two other effects show that a Monte Carlo treatment allowing for cuts and
being able to include all diagrams is indispensable.

The results for Z-pair production again show the effects of cuts and the
inclusion of background diagrams. Although the second energy is tuned in a
way that comparisons to W-pair production may make sense, the energy losses
divided by the total energy are different for these two processes.

\section*{Acknowledgments}
We gratefully acknowledge useful explanations on radiative corrections to
W-pairs by Dr. W. Beenakker and detailed information from Dr. D. Bardin on the
results of ref. \cite{4}.

\begin{table}[ft]
\begin{center}\begin{tabular}{|l|c|c|c|c|c|c|}
\hline \hline
strategy       & $\sigma_0$ & $\sigma_4$ & $\sigma$ &
$\bar{\epsilon}_4$ & $\bar{\epsilon}$ & $\epsilon$ \\
\hline \multicolumn{7}{|c|}{ $\sqrt{s}=$176 GeV}\\ \hline
{\bf WW,f,1}   & .60111 & .50504 & .50490 & 1.168 & 1.162 & --    \\
               & .00032 &        & .00032 &       & 0.002 &       \\ \hline
{\bf WW,s,1,a} & ''     & --     & .50484 & --    & 1.172 & 1.175 \\
               &        &        & .00033 &       & 0.002 & 0.002 \\ \hline
{\bf WW,s,1,b} & ''     & --     & .50175 & --    & 1.167 & 1.170 \\
               &        &        & .00098 &       & 0.006 & 0.006 \\ \hline
{\bf WW,f,2}   & ''     & .50315 &  --    & 1.200 &  --   &  --   \\
               &        &        &        &       &       &       \\ \hline
{\bf WW,s,2}   & ''     & --     & .50258 & --    & 1.178 & 1.181 \\
               &        &        & .00097 &       & 0.006 & 0.006 \\ \hline
{\bf WW,cuts}  & .44651 & --     & .37737 & --    & 1.192 & 1.195 \\
               & .00092 &        & .00091 &       & 0.007 & 0.007 \\ \hline
{\bf all,cuts} & .45011 & --     & .37926 & --    & 1.149 & 1.152 \\
               & .00097 &        & .00095 &       & 0.007 & 0.007 \\
\hline \multicolumn{7}{|c|}{ $\sqrt{s}=$190 GeV}\\ \hline
{\bf WW,f,1}   & .67911 & .60733 & .60750 & 2.108 & 2.092& --    \\
               & .00038 &        & .00039 &       & 0.003 &       \\ \hline
{\bf WW,s,1,a} & ''     & --     & .60706 & --    & 2.111 & 2.118 \\
               &        &        & .00039 &       & 0.003 & 0.003 \\ \hline
{\bf WW,s,1,b} & ''     & --     & .60330 & --    & 2.098 & 2.106 \\
               &        &        & .00117 &       & 0.010 & 0.010 \\ \hline
{\bf WW,f,2}   & ''     & .60696 & --     & 2.171 & --    & --    \\
               &        &        &        &       &       &       \\ \hline
{\bf WW,s,2}   & ''     & --     & .60630 & --    & 2.124 & 2.132 \\
               &        &        & .00117 &       & 0.010 & 0.010 \\ \hline
{\bf WW,cuts}  & .48289 & --     & .43124 & --    & 2.188 & 2.196 \\
               & .00107 &        & .00106 &       & 0.013 & 0.013 \\ \hline
{\bf all,cuts} & .49316 & --     & .44164 & --    & 2.136 & 2.144 \\
               & .00115 &        & .00114 &       & 0.013 & 0.013 \\
\hline \hline
\end{tabular}\end{center}
\caption[.]{Results on radiatively corrected cross sections and average
energy losses, under various calculational strategies, for the process
$e^+ e^- \to e^- \bar{\nu_e} u \bar{d}$. The second line
in each entry is the estimated Monte Carlo error.}
\end{table}
\begin{table}[ft]
\begin{center}\begin{tabular}{|l|l|l|l|l|}
\hline \hline
strategy       & $~~~~~~\sigma_0$ & $~~~~~~\sigma$ &
$~~~\bar{\epsilon}$ & $~~~\epsilon$ \\
\hline \multicolumn{5}{|c|}{ $\sqrt{s}=$190 GeV}\\ \hline
{\bf ZZ,s,2}   &$.70549~10^{-2} $&$.54632~10^{-2} $ &0.744 &0.745 \\
               &$.00040         $&$.00060         $ &0.002 &0.002 \\ \hline
{\bf ZZ,cuts}  &$.52192~10^{-2} $&$.40742~10^{-2} $ &0.732 &0.734 \\
               &$.00060         $&$.00065         $ &0.003 &0.003 \\ \hline
{\bf all,cuts} &$.92509~10^{-2} $&$.78393~10^{-2} $ &1.929 &1.940 \\
               &$.00182         $&$.00182         $ &0.016 &0.016 \\
\hline \multicolumn{5}{|c|}{ $\sqrt{s}=$215.942 GeV}\\ \hline
{\bf ZZ,s,2}   &$.93245~10^{-2} $&$.83903~10^{-2} $ &2.614 &2.624 \\
               &$.00044         $&$.00067         $ &0.008 &0.008 \\ \hline
{\bf ZZ,cuts}  &$.62890~10^{-2} $&$.57262~10^{-2} $ &2.699 &2.709 \\
               &$.00079         $&$.00084         $ &0.011 &0.011 \\ \hline
{\bf all,cuts} &$.99837~10^{-2} $&$.92763~10^{-2} $ &3.503 &3.526 \\
               &$.00194         $&$.00198         $ &0.021 &0.021 \\
\hline \hline
\end{tabular}\end{center}
\caption[.]{Results on radiatively corrected cross sections and average
energy losses, under various calculational strategies, for the process
$e^+ e^- \to e^+ e^- u \bar{u}$. The second line in each entry is the
estimated Monte Carlo error.}
\end{table}

\end{document}